\documentstyle[osa,manuscript,psfig]{revtex} 

\begin{document}
\titlepage

\title{ 
Inelastic diffraction and color-singlet gluon-clusters 
in high-energy hadron-hadron\\
and lepton-hadron collisions}

\author{Meng Ta-chung, R. Rittel and Zhang Yang\\
{\it Institut f\"ur Theoretische Physik,
Freie Universit\"at Berlin,
14195 Berlin, Germany}\protect\\
{\small e-mail: meng@physik.fu-berlin.de; 
        rittel@physik.fu-berlin.de; 
        zhangy@itp.ac.cn}}
\maketitle

\begin{abstract}
It is proposed, that ``the colorless objects'' which manifest
themselves in large-rapidity-gap events are color-singlet
gluon-clusters due to self-organized criticality (SOC), and that
optical-geometrical concepts and methods are useful in examing the
space-time properties of such objects. A simple analytical  
expression for the $t$-dependence of the inelastic single diffractive 
cross section $d\sigma/dt$ ($t$ is the four-momentum transfer squared)
 is derived. Comparison with the existing data and predictions for
future experiments are presented. 
The main differences and similarities between 
the  SOC-approach and the ``Partons in the Pomeron 
(Pomeron and Reggeon)''-approach are discussed.
\end{abstract}


\newpage

Diffraction in Optics can be used as an instrument to 
determine the unknown wavelengths of  incident waves from the known 
geometrical structures of scatterers and vice versa. This has been 
pointed out and 
demonstrated by von Laue and his collaborators in their celebrated paper 
\cite{Laue} eighty-five  years ago. Based on this idea, 
a series of experiments\cite{nuclei} have been performed in the 1950's and 
1960's to measure the sizes of various nuclei by using hadron-beams 
(as hadronic waves) where particle-accelerators have been used as
``super microscopes'' \cite{microscope}. Theoretically,
the idea of using optical and/or geometrical analogies to describe
high-energy hadron-nucleus and hadron-hadron collisions at small
scattering angles has been discussed by many authors\cite{OpticalModels} 
many years
ago. It is shown\cite{OpticalModels} in particular 
that this approach is very useful in
describing 
hadron-hadron
elastic scattering.

Since the recent observation\cite{LRGDiscovery} 
of large-rapidity-gap (LRG) events in 
deep-inelastic electron-proton scattering in the small-$x_B$ region
($x_B < 10^{-2}$, say) -- a kinematical region where soft gluons 
dominate \cite{GluonDominance}, much attention \cite{Gallo}
has been attracted by inelastic diffractive scattering processes in
lepton- and hadron-induced high-energy
collisions\cite{LRGDiscovery,Gallo,DISdata,Breitweg,ppdata,ppbardata}.
The vast interest in this kind of events is mainly due to the fact
that the occurrence of LRG's strongly suggests the existence of certain 
``colorless object(s)'', the ``exchange'' of which plays a dominating
role in such processes.
The ``colorless object'' which carries the quantum-numbers 
of vacuum has been given
\cite{LRGDiscovery,Gallo,DISdata,Breitweg,ppdata,ppbardata,Ingelman,Sterman} 
different names (Pomeron, Reggeon etc.), and it has been suggested
\cite{Ingelman,Sterman} that it (they) should be hadron-like
and thus should have hadron-like fluxes and hadron-like structure-functions.
This class of
inelastic processes are often called
\cite{LRGDiscovery,Gallo,DISdata,Breitweg,ppdata,ppbardata,Ingelman,Sterman} 
``diffractive scattering processes'', because the same kind of
``colorless object'' also plays a significant role in elastic
hadron-hadron scattering, and
the differential cross-section data of the latter 
exhibit diffraction patterns similar to those observed in Optics.

What {\em are} such ``colorless objects''? Can their occurrence, their
properties, and their effects be understood in terms of 
QCD? What is the relationship between such
``colorless objects'' and inelastic diffraction in Optics? 
Can optical-geometrical concepts and methods be used 
to describe inelastic diffractive scattering
processes 
at large values of invariant momentum transfer,
$|t|\ge 0.2\,\mbox{GeV}^2$, say?
In order to 
answer these questions, it seems useful to recall and/or to note 
the following:

(A) A number of experimental 
facts\cite{GluonDominance,Gallo,DISdata,Breitweg} and theoretical 
arguments \cite{F2D3Introduction} suggest that the ``colorless
objects'' which the beam-particles encounter in LRG events are
color-singlet gluon-clusters.

(B) The characteristic properties of the gluons -- especially the
local gluon-gluon coupling prescribed by the QCD-Lagrangian, the
confinement, and the non-conservation of gluon-numbers -- strongly
suggest that systems of interacting soft gluons are
{\em open}, {\em dynamical}, {\em complex} systems 
with {\em many degrees of freedom}, and
that such systems are in general {\em far from equilibrium}.
This means in particular that, since soft gluons
can be emitted and/or absorbed at any time, 
and everywhere in such a system, it is neither meaningful nor 
possible to specify the number of
gluons or the amount of 
energy associated with the system.

(C) It has been pointed out by Bak, Tang and Wiesenfeld
(BTW)\cite{OriginalSOC} 
that a wide class of open complex systems 
evolve into self-organized critical (SOC) states;
and local perturbations
of such critical states may propagate 
like avalanches
caused by
domino effects
over all length scales.
Such a long-range correlation effect 
eventually terminates after a total time $T$, 
having reached a final amount of dissipative
energy, and having effected a total spatial extension $S$. 
The quantity $S$ is called by BTW\cite{OriginalSOC}
``the size'', 
and the quantity $T$ ``the lifetime'' of the ``avalanche''
and/or the ``clusters''. It is 
observed\cite{OriginalSOC,ReviewSOC} that
there are many such open dynamical complex systems in the macroscopic
world, and that
the distributions $D_S$ of $S$, and the distribution
$D_T$ of $T$ of such BTW-avalanches/clusters
obey power laws:
$D_S(S)\propto S^{-\mu}$, and $D_T(T)\propto T^{-\nu}$,
where $\mu$ and $\nu$ are positive real constants. 
Such characteristic behaviors are known
\cite{OriginalSOC,ReviewSOC} as ``the fingerprints'' of
SOC.  Having this, and the characteristic
features of the gluons mentioned in (B), in mind, we are naturally led 
to the question: Can SOC and thus BTW-avalanches also exist in 
microscopic systems -- at the level of quarks and gluons?    
Can SOC be the dynamical origin of color-singlet gluon-clusters which
play the dominating role in inelastic diffractive scattering
processes?

(D) 
In order to answer these questions 
we performed a systematic analysis \cite{SOCI} of the data
\cite{Gallo,DISdata}  
for diffractive DIS where we made
use of the following:
(i) For a color-singlet gluon-cluster
$c_0^\star$ to be a BTW-avalanche, its spatial size 
$S$ has
to be directly proportional to the dissipative energy, and
the latter is proportional to $x_P$ 
which is the energy fraction carried by
$c_0^\star$.
Hence $D_S(S)$ is expected to be related to
the $x_P$-distribution of the $c_0^\star$'s in
LRG events of DIS.
(ii) In such events, $x_P$ and the ``diffractive structure function''
$F_2^{D(3)}$ are measurable
quantities. This, together with the observation that the 
impulse-approximation is applicable to $c_0^\star$'s
(Since the interactions between the struck $c_0^\star$ and any other
neighboring color-singlets are expected to be similar to those between 
hadrons, the former can 
also be considered as Van der Waal's
type of interactions.), implies that
$D_S(S)$ can be extracted
 in a similar manner as the $x_B$-distribution of 
valence quarks
in normal events.
Furthermore, if the $c_0^\star$'s are indeed due to SOC,
there should be $c_0^\star$'s of all lifetimes ($T$'s). This means,
for a given interaction-time $\tau_{\mbox{\tiny int}}$, there are
always $c_0^\star$'s whose $T$'s are longer than 
$\tau_{\mbox{\tiny int}}$.
(iii) The interaction-time $\tau_{\mbox{\tiny int}}$ can be 
estimated by making use of the
uncertainty principle; to be more precise
(cf. Eq.(10) of Ref.[\ref{SOCI}]), 
for fixed
$|\vec P|$ and $Q^2$ values, $\tau_{\mbox{\tiny int}}\propto x_B$ for $x_B\ll
1$. Hence, in the small-$x_B$ region, information about the
lifetime-distribution can be obtained by examining the
$x_B$-dependence of $F_2^{D(3)}$.
The results of the analysis
can be summarized as follows: The size- and the 
lifetime-distributions are indeed of the form $D_S(S)\propto S^{-\mu}$ and
$D_T(T)\propto T^{-\nu}$ where $\mu\approx\nu\approx 2$. 
Furthermore, for obvious reasons,
$D_S(S)$ and $D_T(T)$ are expected to retain their power-law behaviors with
exactly the same exponent in all Lorenz-frames.
Comparisons between
these results and those obtained for inelastic diffractive
$\gamma p$\cite{Breitweg}, $pp$\cite{ppdata}, 
and $\bar{p}p$\cite{ppbardata} processes
have also been made\cite{SOCI}, and the following picture emerges: 
In an inelastic diffractive scattering process off
proton the beam-particle encounters one of the color-singlet
gluon-clusters which in general exist partly inside and 
partly outside the confinement region
of the proton. The size- and the lifetime-distributions of the
clusters exhibit universal power-law behaviors, 
which imply in particular
that such gluon-clusters are {\em not}
hadron-like in the sense that they 
have {\em neither} a typical size, {\em nor} a typical
lifetime. Furthermore, the fact\cite{SOCI}
that the data\cite{DISdata} {\em cannot} accommodate the simple factorization 
assumption\cite{Ingelman} in which a universal pomeron-flux with
a unique hadron-like pomeron structure function exist, 
gives further support to the proposed SOC-picture
because a BTW-cluster {\em cannot} have a universal static structure.
With these characteristic properties
of the colorless objects in mind, we see\cite{Sterman} an overlap
between the SOC-picture and  the partonic picture for
Pomeron and/or 
Pomeron and Reggeon\cite{Ingelman,Sterman} in which, beside
the Pomeron, exchange of (in general an infinite number of)
subleading Regge-trajectories are possible. In 
fact, it has been reported\cite{DISdata} that very good agreement 
can be achieved between the data\cite{DISdata} and this kind of models.
Hence, in order to differentiate between the two approaches, it is not 
only useful but also necessary to examine the corresponding
predictions for the dependence on the invariant momentum-transfer $t$.

(E) In Optics,  Frauenhofer diffraction can be observed when
the wavelength of the light is less than the linear dimension 
of the scatterer, and when  the light-source and 
the detecting screen are very far from the scatterer. 
The incident light-rays are considered as plane waves
(wave vector $\vec k$, frequency $\omega\equiv |\vec k|$).
After the scattering, the field originating from the scatterer  
can be written as the
product of $R^{-1}\exp{(i|\vec{k'}|R)}$ and 
the scattering amplitude
$f(\vec k,\vec{k'})$. Here,
$\vec {k'}$ is the wave vector of the scattered light 
in the direction of observation, $\omega^{\prime} \equiv |\vec {k'}|$ 
is the corresponding frequency, $R$ is the distance between the 
scatterer and the observation point.
By choosing a coordinate system in which the $z$-axis coincides with $\vec k$, 
$f(\vec k,\vec{k'})$ can be expressed\cite{Landau} as
\begin{eqnarray}\label{ffourier} 
f(\vec q) & = & 
(2\pi)^{-2} \int\mbox{\hspace*{-0.3cm}}\int_{\Sigma}^{} 
d^2\vec{b}\ \alpha(\vec{b})\,
e^{-i\vec{q} \cdot \vec{b}}\mbox{\ .}
\end{eqnarray}
Here,  $\vec q\equiv \vec {k'} - \vec k$ determines 
the change in wave vector due to diffraction; $\vec{b}$ is the 
impact parameter, $\alpha(\vec{b})$ 
is the corresponding amplitude in the two-dimensional
$\vec b$-space (here, the $xy$-plane), and the
integration extends over the region $\Sigma$ in which
$\alpha(\vec{b})$ is different from zero. 
It is well-known  that elastic scattering at small angles can be
deduced\cite{Landau} from this equation under the additional
condition
$|\vec{k}^\prime|=|\vec{k}|=\omega^\prime=\omega$, 
where geometry dictates that
$\vec{q}$ is approximately perpendicular to $\vec{k}$ and to 
$\vec{k}^\prime$.
In such cases, $\vec{q}\approx\vec{q}_{\perp}$
where $\vec{q}_{\perp}$ stands for its projection on
the
$xy$-plane of the chosen coordinate system.
We note, the general case, in which $\vec{k}^{\prime}\neq\vec{k}$ and
$\omega^{\prime}\neq\omega$, corresponds to inelastic scattering.
for which we can write
\begin{equation}
\label{finelfourier}
f_{\rm inel}(\vec{q}_\perp)= (2\pi)^{-2}
\int\mbox{\hspace*{-0.3cm}}\int_{\Sigma}^{} 
d^2\vec b\ \alpha_{\rm
inel}(\vec b)e^{-i\vec{q}_\perp \cdot\vec b}\mbox{\,.}
\end{equation}
Here both $\alpha_{\mbox{\tiny inel}}(\vec b)$ and 
$f_{\mbox{\tiny inel}}(\vec{q}_\perp)$ in general depend on 
the transfer of energy, and/or longitudinal momentum.
Furthermore, if the scatterer is 
symmetric with respect to the scattering axis, 
Eq.(\ref{finelfourier}) can also  be expressed by
using an integral representation for $J_0$, as
\begin{equation}
\label{fbessel}
f_{\rm inel}(q_\perp)= (2\pi)^{-1}\int_0^\infty b\,db\ \alpha_{\rm
inel}(b)J_0(q_\perp b),
\end{equation}
where $q_\perp$ and $b$ stand for $|\vec {q_\perp}|$ and $|\vec{b}|$ 
respectively. 

Keeping the facts mentioned in (A) -- (E) in mind, we now examine the
scatterer quantitatively in the rest frame of the proton target. 
We  choose a right-handed Cartesian coordinate system with its origin $O$ at 
the center of the proton and the $z$-axis in the direction of the incident 
beam-particle which is considered point-like as it goes through the $xy$-plane 
at point $B\equiv(0,b)$. That is, the incident beam and the center $O$ of the 
proton determine the scattering plane ($yz$-plane) of the collision event, 
where the distance $\overline{OB}$ is the corresponding impact parameter 
$b$. Since we are dealing with inelastic scattering (where the momentum 
transfer  also in the longitudinal direction can be large) it is possible 
to envisage that the scattering 
takes place at one point in space, and
effectively\cite{Footnote} 
at the point $B$, where it meets
colorless gluon-clusters. The latter are BTW-avalanches 
due to SOC
initiated by local 
interactions
in systems of soft gluons. Since gluons 
carry color, the interactions which lead to the formation of
gluon-clusters  must take place 
inside the confinement region of  the proton. This means, while a 
considerable part of the created color-singlet clusters
can be outside
the proton, the location $A$ where such an avalanche is initiated 
{\em must} be {\em inside} the proton. That is, in terms of 
$\overline{OA}\equiv r$, $\overline{AB}\equiv R_A(b)$, and  
proton's radius $r_p$, we have $r\le r_p$ and 
$[R_A(b)]^2=b^2+r^2-2b r \cos\angle BOA$. For a given impact parameter
$b$, the relevant quantity
is the mean distance $\langle R_A^2(b)\rangle^{1/2} = ( b^2 + a^2)^{1/2}$,
$a^2\equiv 3/5\,r_p^2$,
which is
obtained by averaging  over all allowed locations of $A$
in the confinement region.
That is, we can model {\em the effect of confinement} in 
avalanche-formation by picturing that all such BTW-avalanches
in particular those
which contribute to 
scattering events characterized by a given $b$
are initiated from an ``effective initial point'' $\langle A_b\rangle$.
Note also that since an avalanche is a dynamical object, it in general 
expands and propagates 
within its lifetime
(in any one of the $4\pi$
directions away from $\langle A_b\rangle$).
It is envisaged that the expansion and propagation of such an
avalanche cannot proceed without limits. This is not only because the
gluon-density is expected to decrease with increasing distance from 
$O$, but also because color-forces between the constituents of 
$c_0^\star$ increase (consistent with confining potentials) when their
distance become larger. That is, those constituents ``going too far''
will be ``pulled back'' to the proton's confinement region.

An explicit expression for
the amplitude $\alpha_{\mbox{\tiny inel}}(b)$ 
in Eq.(\ref{fbessel}),
can be readily written down by taking
the following into account:
($\alpha$) SOC dictates that there are avalanches of all sizes ($S_i$'s) and
that the number-density of avalanches of $S_i$ is
$D_S(S_i)\propto S_i^{-\mu}$ where 
experiments show\cite{SOCI} $\mu\approx 2$.
($\beta$)
Since
the interactions between the struck  
$c_0^\star$ and any other color-singlets
should be of Van der Waals type, 
it can simply be 
``carried away'' by the beam-particle.
Simple geometrical considerations suggest that the chance for an
avalanche of size (i.e. volume) $S_i$ to be hit
(on the plane perpendicular to the incident axis) 
should be proportional to $S_i^{2/3}$.
($\gamma$) Since for a given $b$, the distance
in space between $\langle A_b\rangle$ and
$B\equiv (0,b)$ is 
$(b^2+a^2)^{1/2}$, the number of avalanches which pass a unit area 
on the shell of radius $(b^2+a^2)^{1/2}$ centered at $\langle A_b\rangle$
is proportional to $(b^2+a^2)^{-1}$, provided that (due to  causality)
the lifetimes ($T$'s) of these avalanches are 
not shorter than
$\tau_{\mbox{\tiny min}}(b)$. The latter is the time interval for an
 avalanche to travel (with constant velocity, say)
from  $\langle A_b\rangle$ to $B$ that is
$\tau_{\mbox{\tiny min}}(b)\propto (b^2+a^2)^{1/2}$, and 
only those avalanches
having lifetimes
$T\ge\tau_{\mbox{\tiny min}}(b)$ can contribute to such an
event. Now, since avalanches are due to SOC 
and the chance for an avalanche
of lifetime  
$T$ to exist is 
$D_T(T) \propto T^{-\nu}$ where\cite{SOCI}  $\nu\approx 2$, 
the proper fraction can be obtained
by integrating 
$ T^{-2}$ over $T$ from $\tau_{\mbox{\tiny min}}(b)$ to infinity,
which is $(b^2+a^2)^{-1/2}$.

Hence, 
for inelastic diffractive scattering processes
in which the beam-particles encounter avalanches of size $S_i$,
the amplitude
which we denote 
by $\alpha_{\mbox{\tiny inel}}(b|S_i)$
can be obtained from the square-root of
$D_S(S_i)$
mentioned in
($\alpha$), and by
taking the weighting factors mentioned in ($\beta$) and ($\gamma$) into
account. The result is:
\begin{eqnarray}\label{alpha}
\alpha_{\rm inel}(b|S_i) &=\mbox{const.}\,S_i^{-1/3}(b^2+a^2)^{-3/2}.
\end{eqnarray}
By inserting this 
in Eq.(\ref{fbessel}), we obtain
the corresponding amplitude in $\vec{q}_\perp$-space:
$f_{\rm inel}(q_\perp |S_i)$.
Here, $q_\perp =|\vec{q}_\perp|\approx\sqrt{|t|}$
is the corresponding
momentum-transfer. 
The integration can be carried out
analytically\cite{IntegralTabelle}, and the result 
(after the summation over $i$)
for the inclusive inelastic single diffractive
differential cross-section $d\sigma/dt$ is
\begin{eqnarray}
\label{sigma}
d\sigma/dt & = &
C \exp{(-2a\sqrt{|t|})}\mbox{\, .}
\end{eqnarray}
Here $a^2\equiv \frac{3}{5} r^2_p$,
and $C$ is an unknown 
constant. Note that the value
for $a$ is
{\em the same} for different
incident energies, and for different
projectiles: $\gamma^\star$, $\gamma$, $p$, $\bar{p}$, etc.

To compare this result with experiments,
we calculate $a$ by using its definition
$a^2\equiv \frac{3}{5} r^2_p$ and the experimental value\cite{Halzen}
for $r_p$ which is $0.81\,\mbox{fm}$. We note, while 
``the slope'' $2a$ 
depends only on the radius of the target-hadron,
the normalization constant $C$ is expected to
be different for different 
projectiles. The comparisons are
shown in Figs.1 and 2. 

A few remarks should be made in this
connection:
(i) The $d\sigma/dt$-data for $pp$, $\bar{p}p$, $\gamma p$ and 
$\gamma^\star p$ reactions are consistent with {\em no}
energy ($\sqrt{s}$ and/or $W$)-dependence. In particular, in contrast
to elastic $pp$-scattering, there is no indication of ``shrinkage''.
(ii) The figures show: In a limited
range of $\sqrt{|t|}$, curves with very much different $t$-behaviors can  
yield rather similar results.
(iii) The existing data show that the coefficients of
$\sqrt{|t|}$,$t$, or $t^2$ in the curves for $d\sigma/dt$ are
consistent with no $M_x$-dependence. This is in particular the case
for $\gamma p$-reactions (see in this connection the second paper of
Ref.[\ref{Breitweg}]).
(iv) A simple analytic expression for
$d^2\sigma/dt\,dx_P$ can be readily obtained by taking 
$S_i \propto x_P$ into account in Eq.(\ref{alpha}) and by integrating over
$b$. The result which is shown to be in good agreement
 with the existing data for $pp$-
and $\bar{p}p$-reactions
will be presented in Ref.[\ref{SOCIII}].
Evidently, further
experiments, especially  measurements of
$d\sigma/dt$ and $d^2\sigma/dt\,d(M_x^2/W^2)$ for
$\gamma p \rightarrow X p$ at larger $|t|$-values
would be helpful.

We thank T. T. Chou for correspondence, 
K. Tabelow and W. Zhu for discussions, and FNK der FU Berlin
for financial support. Y. Zhang thanks Alexander 
von Humboldt Stiftung for the fellowship granted to him.

\begin{figure}[hbt]
\caption{
The $d\sigma/dt$ data taken from 
Refs.[\ref{ppdata},\ref{ppbardata}]
are plotted against $\sqrt{|t|}$ in the measured kinematical range.
The solid line is our result.
The dashed and dot-dashed lines show the conventional fits
$d\sigma/dt \propto \exp{[B t + C t^2]}$.
The former is the UA4-fit to their data [\ref{ppbardata}]
with $B= 8.0\pm 0.1\,\mbox{GeV}^{-2}$ and $C= 2.3\pm 0.1\,\mbox{GeV}^{-4}$.
The latter is a fit to the same expression for
all the data points in this figure, where 
$B=  5.7\pm 0.1\,\mbox{GeV}^{-2} $ and $C= 0.8\pm 0.1\,\mbox{GeV}^{-4}$.}
\end{figure}

\begin{figure}[hbt]
\caption{
The  $d\sigma /d|t|$ data from 
Refs.[\ref{Gallo},\ref{Breitweg}]  
for $\gamma^{\star}$  and $\gamma$ 
induced reactions
are plotted against $\sqrt{|t|}$. 
They are shown as circles
and squares respectively.
Here, the empty circles and squares
are the data from 
(the transparencies of) Ref.[\ref{Gallo}]
while the solid ones are those from 
Refs.[\ref{Breitweg}].
The solid lines
stand for our result.
The dashed and dotted lines are fits given in
Refs.[\ref{Gallo},\ref{Breitweg}]: 
$d\sigma/dt \propto\exp{(- b |t|)}$,
where $b = 7.3\pm 0.9\pm 1.0\,\mbox{GeV}^{-2}$
and $b= 6.8\pm 0.9 +1.2/-1.1\,\mbox{GeV}^{-2}$.}
\end{figure}

\vspace*{-1cm}
\hspace*{-1cm}
\psfig{figure=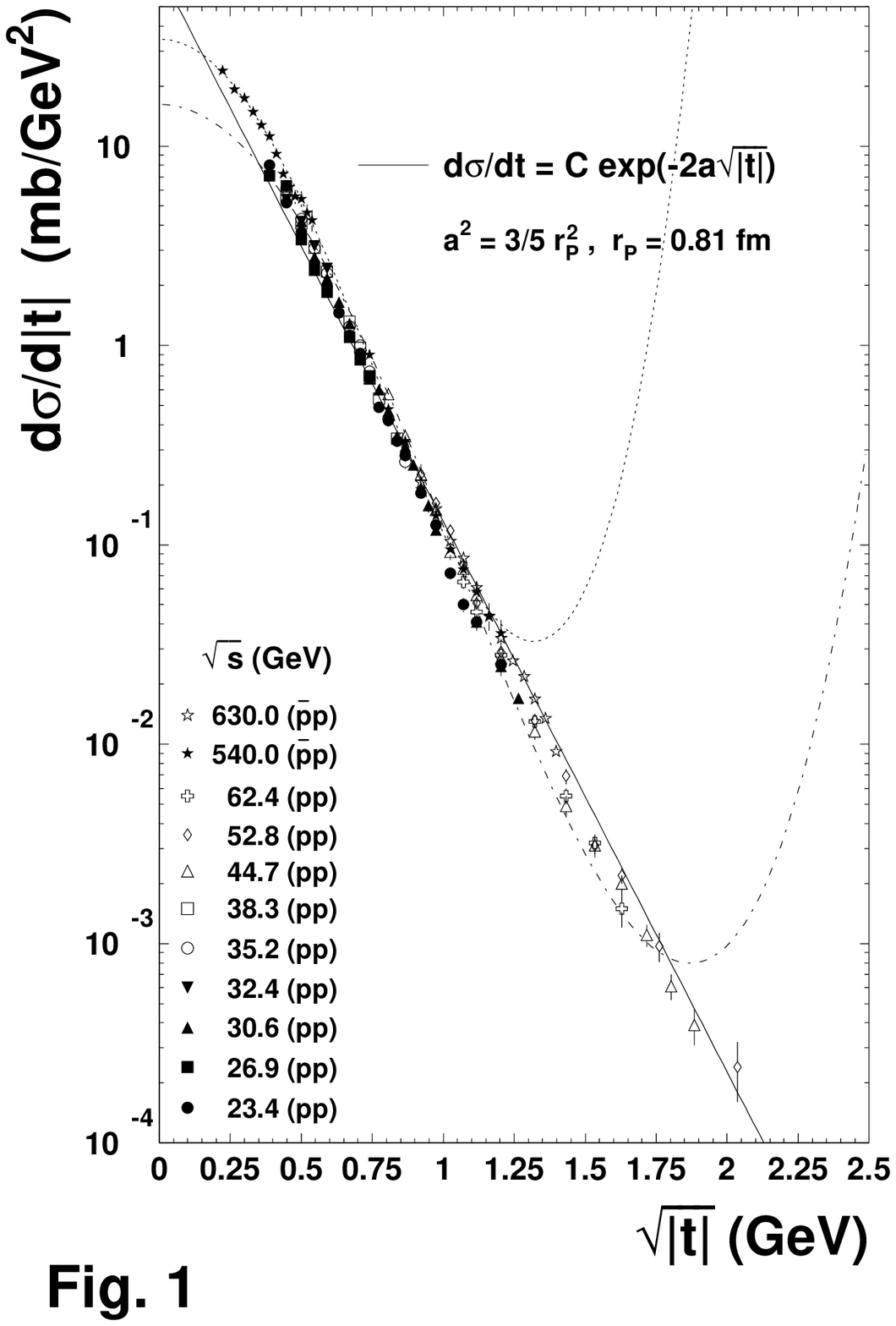,height=13.13cm}

\vspace*{-2cm}
\hspace*{-2.5cm}
\psfig{figure=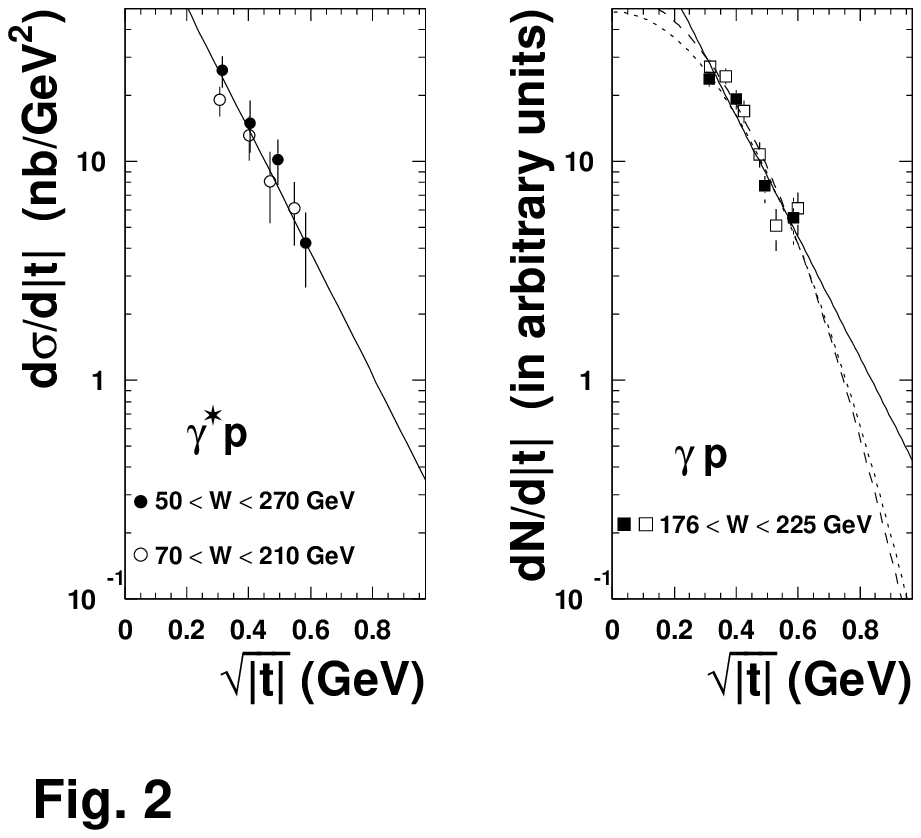,height=10cm}


\begin{thebibliography}{99}
\bibitem{Footnote}
\label{Footnote}
This is the point from where the main contribution comes.
Contributions from points outside the $xy$-plane, and effects caused by 
the environment on the propagating avalanche, can be, and have been,
taken into account in a more elaborated modeling
procedure\cite{SOCIII}. The final result turns out to be the same.
\end{thebibliography}
\end{document}